\documentclass[12pt]{article}
\usepackage{amsmath}
\usepackage{amssymb}
\usepackage[english]{babel}
\usepackage{graphicx}
\usepackage{subfigure}
\usepackage{indentfirst}
\usepackage{hep}
\usepackage{mathrsfs}
\usepackage{dsfont}
\usepackage{setspace}
\usepackage[labelfont=bf,font=footnotesize,labelsep=period]{caption}
\usepackage{xcolor}
\usepackage{float}
\usepackage{hyperref}
\hypersetup{
colorlinks=true,
linkcolor=blue,
citecolor=blue,
}
\usepackage[capitalize]{cleveref}
\usepackage[numbers,sort&compress]{natbib}

\begin{document}
\doublespacing

\title{Leggett-type inequalities for testing nonlocal realism in multipartite systems\thanks{Corresponding author: qiaocf@ucas.ac.cn}}

\author{
Abdul Sattar Khan$^{1,2,3}$, Ma-Cheng Yang$^{1}$, and Cong-Feng Qiao$^{1,4,*}$ \\ [0.2cm]
\footnotesize{$^{1}$School of Physical Sciences, University of Chinese Academy of Sciences, Beijing 100049, China}\\
\footnotesize{$^{2}$School of Physics, Shandong University, Jinan, China}\\
\footnotesize{$^{3}$Wilczek Quantum Center, School of Physics and Astronomy, Shanghai Jiao Tong University, Shanghai 200240, China}\\
\footnotesize{$^{4}$Key Laboratory of Vacuum Physics, CAS, Beijing 100049, China}
}
\date{\today}

\maketitle

\begin{abstract}
Nonlocal realism represents the last classical cornerstone in conflict with quantum theory, and has been shown to be largely untenable in bipartite systems [Nature \textbf{446}, 871 (2007); Nature Physics \textbf{4}, 681 (2008)]. We extend the Leggett-type nonlocal realistic model to arbitrary $N$-partite systems with polarizer settings, and derive rigorous inequalities that distinguish quantum predictions from nonlocal realistic theories. The derivation yields a fundamental double inequality that is universally valid, from which the Leggett-type bounds follow under specific measurement settings. As an illustration, we demonstrate quantum violations of these inequalities for Greenberger--Horne--Zeilinger (GHZ) states, with maximal violation $2(\sqrt{5}+1)$. Our results show that nonlocal realism in multipartite systems is experimentally testable.
\end{abstract}

\newpage

\section{Introduction}

Realism is the philosophical view that an external reality exists and has definite properties independent of observation \cite{Clauser1978,Chiao1999,Groeblacher2007,Leggett2008,Karakostas2012}. Einstein, Podolsky, and Rosen (EPR) first argued that the quantum wave function provides an incomplete description of physical reality \cite{Einstein1935}. Building on the EPR argument using the example proposed by Bohm and Aharonov \cite{Bohm1957}, Bell constructed inequalities based on the assumptions of local realism \cite{Bell1964,Clauser1978,Aspect1999,Norsen2009,Gill2014,Hensen2015,Handsteiner2017,Li2018,Abellan2018,Blasiak2021,Qian2021,Bhowmik2021,Takubo2021,Ruzbehani2021}. Quantum theory predicts violations of these inequalities for certain measurements on entangled particles \cite{Clauser1969,Clauser1974,Aspect2015}, implying that at least one of the underlying assumptions — locality or realism — must fail.

In 2003, Leggett introduced a class of nonlocal hidden-variable theories, termed crypto-nonlocal, motivated by the observation that the incompatibility between objective local theories and quantum mechanics has less to do with locality than with objectivity (realism) \cite{Leggett2003}. He formulated an incompatibility theorem between nonlocal realism and quantum physics in the form of the Leggett inequality.

Soon afterward, following Leggett's approach, Groeblacher et al. \cite{Groeblacher2007} conducted an experimental test of nonlocal realism, ruling out a broad class of nonlocal hidden-variable theories against quantum theory. Experimental falsifications of the model were reported without additional assumptions soon after \cite{Paterek2007,Branciard2007}. For critiques and replies, see Refs. \cite{Laudisa2008,Laudisa2014,Egg2013}.

In 2008, the Leggett model was generalized using two main assumptions: no-signaling and that locally everything happens as in pure quantum states \cite{Branciard2008}. A new approach with stronger inequalities was presented, and the simplest of these inequalities was shown to be violated by quantum theory experimentally. For discussions on whether hidden-variable models for quantum theory can have local parts, see Refs. \cite{Colbeck2008,Larsson2009}.

Tests of the Leggett inequality with different systems \cite{Paternostro2010,Romero2010,Hasegawa2012,Cardano2013} were in agreement with quantum physics. Leggett-type inequalities without geometric constraints \cite{Rai2011}, with relaxed measurement independence \cite{Zela2013}, and with different interpretations \cite{Lorenzo2013} have also been studied. Recent work includes falsifications with solid-state spins \cite{Huang2020}, violations in high-energy physics \cite{Khan2020,Shi2020}, and alternative nonlocal models \cite{Suarez1997,Zbinden2001,Stefanov2002}.

All previous efforts have focused on two-party scenarios. Going from two parties to three makes the situation more complicated and interesting. Much work has been devoted to multipartite Bell nonlocality \cite{Svetlichny1987,Mitchell2004,Gallego2012,Bancal2013}, but multipartite Leggett-type tests remain largely unexplored, with only one prior attempt \cite{Deng2011}. Given the advantage of having exact measurement settings to test experimentally, we here extend Leggett-type inequalities to arbitrary $N$-partite systems.

The paper is organized as follows. In Sec.~II, we introduce the $N$-partite Leggett model. In Sec.~III, we present the rigorous derivation of the Leggett-type inequalities. In Sec.~IV, we demonstrate violations for the tripartite GHZ state. We conclude in Sec.~V.

\section{Multiparty Leggett Model}

We consider qubits encoded in photon polarization, as in the bipartite case \cite{Leggett2003,Groeblacher2007,Branciard2008}. The observable $\mathcal{X}$ is parameterized by unit vectors $\vec{n}$ on the Poincar\'e sphere, i.e., $\mathcal{X} = \vec{\sigma}\cdot \vec{n}$, where $\sigma_{\mu}$ are the Pauli matrices. We use the notation:
\begin{align}
x &:= x_1,x_2,\ldots,x_N, \\
\lambda &:= \vec{u}_1,\vec{u}_2,\ldots,\vec{u}_N, \\
\mathcal{X} &:= \mathcal{X}^{(1)},\mathcal{X}^{(2)},\ldots,\mathcal{X}^{(N)}.
\end{align}
Here, $x_i = \{\pm 1\}$ are the binary outcomes for the $i$th particle, and $\vec{u}_i$ is its polarization direction.

\begin{figure}[H]\centering
\scalebox{0.20}{\includegraphics{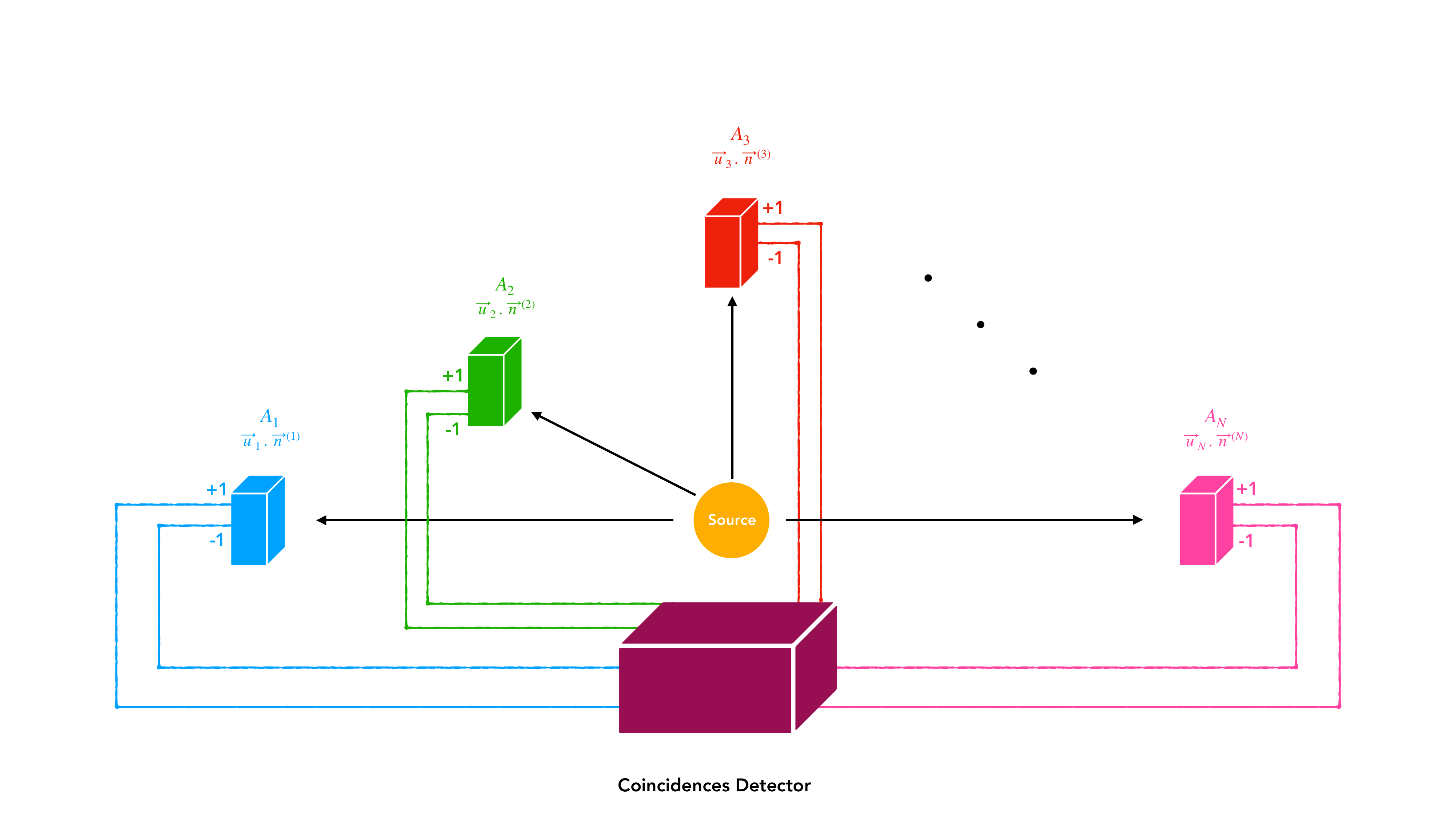}}
\caption{Visualization of the $N$-partite experimental setup. A common source emits $N$ photons in polarization product states $\bigotimes_{i=1}^{N}|\vec{u}_i\rangle$, such that each photon is in a definite polarization state, and the local expectation obeys Malus' law.}
\label{fig:Npartite-experimental-setup}
\end{figure}

The source emits $N$ photons described by polarization product states $\bigotimes_{i=1}^{N}|\vec{u}_i\rangle$. The local expectation satisfies Malus' law:
\begin{align}
\langle x_i\rangle_{\lambda} = \langle \vec{u}_i|\mathcal{X}^{(i)}|\vec{u}_i\rangle = \vec{u}_i\cdot \vec{n}^{(i)}.
\end{align}

The $N$-partite joint probability is:
\begin{align}
\mathcal{P}(x|\mathcal{X}) = \int d\lambda\,\rho(\lambda)\mathcal{P}_{\lambda}(x|\mathcal{X}),
\end{align}
where
\begin{align}
\mathcal{P}_{\lambda}(x|\mathcal{X}) &= \frac{1}{2^N}\Bigg[1 + \sum_{i=1}^{N}x_i\mathcal{C}_{\lambda}^{(i)}(\mathcal{X}) + \sum_{i_1<i_2}x_{i_1}x_{i_2}\mathcal{C}_{\lambda}^{(i_1,i_2)}(\mathcal{X}) \notag \\
&\quad + \cdots + x_1x_2\cdots x_N\mathcal{C}_{\lambda}^{(1,2,\ldots,N)}(\mathcal{X})\Bigg].
\end{align}

The marginals and correlation coefficients are defined as:
\begin{align}
&\mathcal{C}_{\lambda}^{(i)}(\mathcal{X}) = \sum_{x_1,\ldots,x_N} x_i\,\mathcal{P}_{\lambda}(x|\mathcal{X}), \\
&\mathcal{C}_{\lambda}^{(1,\ldots,N)}(\mathcal{X}) = \sum_{x_1,\ldots,x_N} x_1\cdots x_N\,\mathcal{P}_{\lambda}(x|\mathcal{X}), \\
&\mathcal{C}_{\lambda}^{(i_1,\ldots,i_j)}(\mathcal{X}) = \sum_{x_1,\ldots,x_N} x_{i_1}\cdots x_{i_j}\,\mathcal{P}_{\lambda}(x|\mathcal{X}).
\end{align}

The no-signaling condition gives:
\begin{align}
&\mathcal{C}_{\lambda}^{(i)}(\mathcal{X}) = \mathcal{C}_{\lambda}^{(i)}(\mathcal{X}^{(i)}), \\
&\mathcal{C}_{\lambda}^{(i_1,\ldots,i_j)}(\mathcal{X}) = \mathcal{C}_{\lambda}^{(i_1,\ldots,i_j)}(\mathcal{X}^{(i_1)},\ldots,\mathcal{X}^{(i_j)}).
\end{align}

Unlike Bell's model, we impose no constraints on the correlation coefficients, allowing for nonlocal correlations.

\section{Rigorous Derivation of $N$-partite Leggett-type Inequalities}

From the nonnegativity of probabilities and Malus' law, we obtain the fundamental constraint (see the Supplementary Material for details):
\begin{align}
\left\{
\begin{aligned}
&|\mathcal{C}_{\lambda}^{(i)}(\mathcal{X}^{(i)}) \pm \mathcal{C}_{\lambda}^{(i_1,\ldots,i_{N-1})}(\mathcal{X})| \leq 1 \pm \mathcal{C}_{\lambda}^{(1,\ldots,N)}(\mathcal{X}), \\
&|\mathcal{C}_{\lambda}^{(i)}(\mathcal{X}^{(i)'}) \pm \mathcal{C}_{\lambda}^{(i_1,\ldots,i_{N-1})}(\mathcal{X}')| \leq 1 \pm \mathcal{C}_{\lambda}^{(1,\ldots,N)}(\mathcal{X}').
\end{aligned}
\right.
\label{eq:core}
\end{align}

For two measurement settings $\mathcal{X}$ and $\mathcal{X}'$, we define:
\begin{align}
\gamma_{\pm}^{(i)}(\lambda) &= \mathcal{C}_{\lambda}^{(i)}(\mathcal{X}^{(i)}) \pm \mathcal{C}_{\lambda}^{(i)}(\mathcal{X}^{(i)'}), \\
\alpha_{\pm}(\lambda) &= \mathcal{C}_{\lambda}^{(i_1,\ldots,i_{N-1})}(\mathcal{X}) \pm \mathcal{C}_{\lambda}^{(i_1,\ldots,i_{N-1})}(\mathcal{X}'), \\
\beta(\lambda) &= \mathcal{C}_{\lambda}^{(1,\ldots,N)}(\mathcal{X}) + \mathcal{C}_{\lambda}^{(1,\ldots,N)}(\mathcal{X}').
\end{align}

Adding the positive inequalities from Eq.~(\ref{eq:core}) gives:
\begin{align}
\gamma_{\pm}^{(i)}(\lambda) \leq 2 \mp [\alpha_{\pm}(\lambda) - \beta(\lambda)].
\label{eq:gamma_upper}
\end{align}

Adding the negative inequalities gives:
\begin{align}
-\gamma_{\pm}^{(i)}(\lambda) \leq 2 \pm [\alpha_{\pm}(\lambda) + \beta(\lambda)].
\label{eq:gamma_lower}
\end{align}

Equations (\ref{eq:gamma_upper}) and (\ref{eq:gamma_lower}) are the fundamental inequalities. Combining them yields the universally valid double inequality:
\begin{align}
-2 + |\alpha_{\pm}(\lambda) + \beta(\lambda)| \leq \gamma_{\pm}^{(i)}(\lambda) \leq 2 - |\alpha_{\pm}(\lambda) - \beta(\lambda)|.
\label{eq:double}
\end{align}

This double inequality is the central result of our derivation. It is always valid and does not require any additional assumptions.

\medskip
\noindent\textbf{A note on the derivation:} The double inequality in Eq.~(\ref{eq:double}) is the most general and rigorous form of our Leggett-type constraint. It is always valid and follows directly from the axioms. In the specific case where the measurement settings satisfy $(\alpha_{\pm})_k = 0$, the double inequality reduces to the simpler form in Eq.~(26) in the supplemental material. This is the form used for the GHZ state violations. However, we emphasize that the simpler $ \max $ bound is not universally valid for arbitrary correlation functions. A counterexample exists for $\alpha = 2$, $\beta = -1$, where the $ \max $ bound fails. The double inequality in Eq.~(\ref{eq:double}) is the correct general result. All violation results reported in this work are obtained under the specific measurement settings where the simplified form applies, and remain valid.
\medskip

For three $2N$-tuples of settings with the same angle $\theta$, using $\sum_{k=1}^{3}|\vec{u}_i\cdot \vec{e}_k| \geq 1$, we obtain:
\begin{subequations}
\begin{align}
\sum_{k=1}^{3} |\gamma_{\pm,k}^{(i)}| + \sum_{k=1}^{3} \max\left(|(\alpha_{\pm})_k - \beta_k|, |(\alpha_{\pm})_k + \beta_k|\right) \leq 6.
\label{eq:main_ineq}
\end{align}
\end{subequations}

Under the specific measurement settings where $(\alpha_{\pm})_k = 0$, Eq.~(17) in the supplemental material reduces to the simpler form:
\begin{subequations}
\begin{align}
\mathcal{L}_{+} &\equiv \sum_{k=1}^{3}|\beta_k| + 2|\cos(\theta/2)| \leq 6, \\
\mathcal{L}_{-} &\equiv \sum_{k=1}^{3}|\beta_k| + 2|\sin(\theta/2)| \leq 6.
\end{align}
\end{subequations}

These are the Leggett-type inequalities used for the GHZ state violations.

\section{Tripartite Violations for the GHZ State}

\begin{figure}
\centering
\subfigure[Alice's measurement settings]{\includegraphics[width=0.48\linewidth]{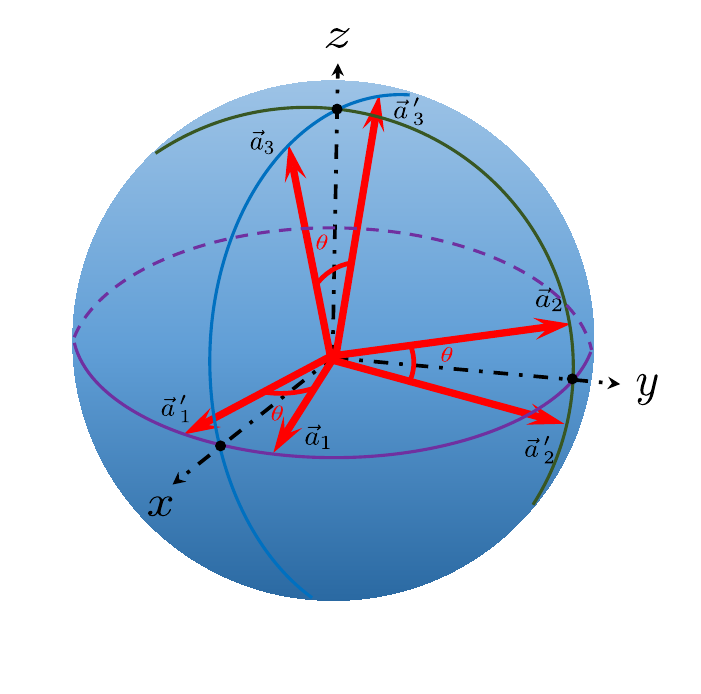}}
\subfigure[Bob and Charlie's measurement settings]{\includegraphics[width=0.48\linewidth]{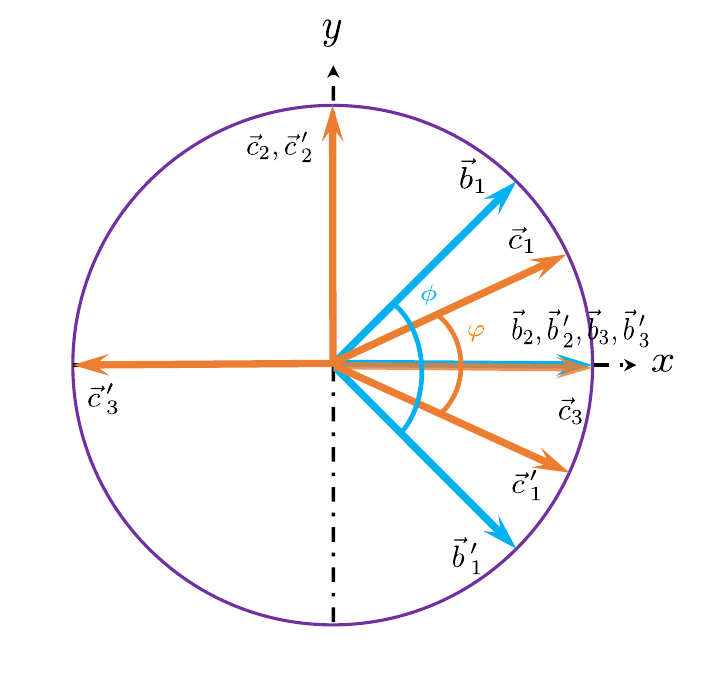}}
\caption{Measurement settings for the tripartite GHZ state.}
\label{fig:measurement_settings}
\end{figure}

For the tripartite GHZ state with the measurement settings in Fig.~\ref{fig:measurement_settings}, we obtain:
\begin{subequations}
\begin{align}
\mathcal{L}_{+}^{\mathrm{GHZ}} &= 2\sqrt{5}\sin(\theta_0 + \theta/2) + 2\left|\cos\frac{\theta + \phi + \varphi}{2}\right|, \label{eq:Lplus} \\
\mathcal{L}_{-}^{\mathrm{GHZ}} &= 2\sqrt{5}\cos(\theta_0 - \theta/2) + 2\left|\cos\frac{\theta + \phi + \varphi}{2}\right|, \label{eq:Lminus}
\end{align}
\end{subequations}
where $\theta_0 = \arctan 2$, $\theta \in [0,\pi]$, and $\phi,\varphi \in [0,2\pi]$.

\begin{figure}[H]\centering
\scalebox{0.90}{\includegraphics{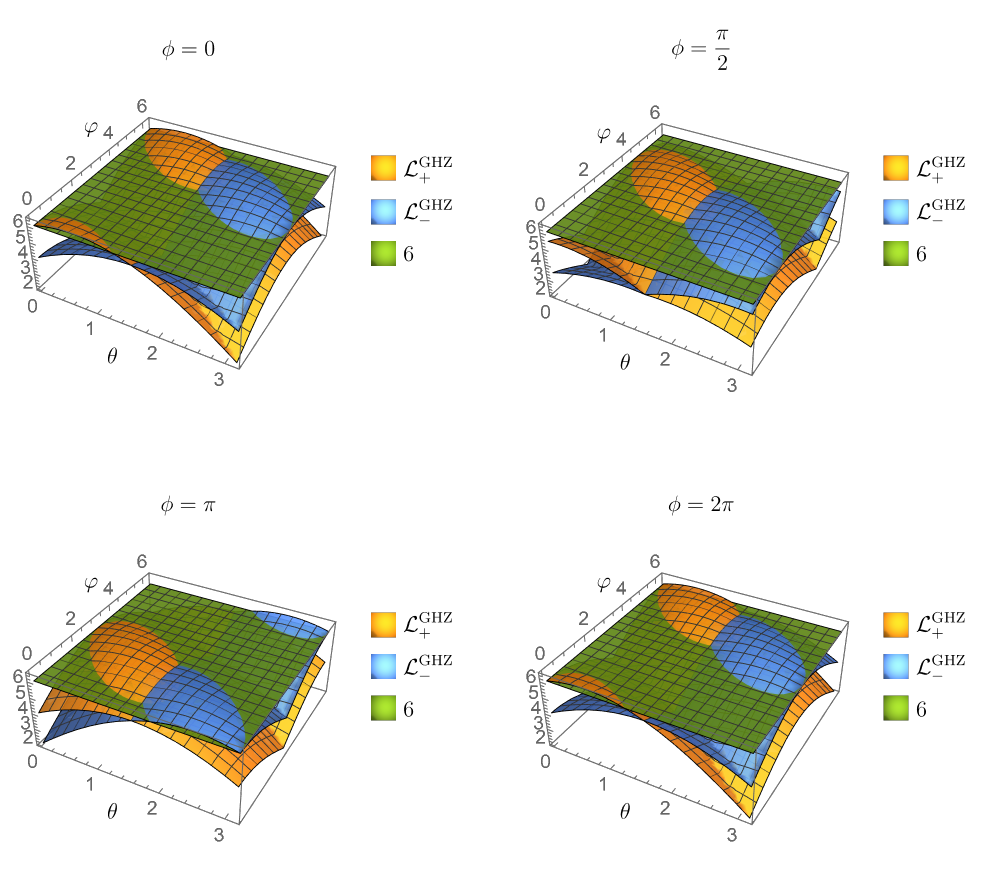}}
\caption{Quantum violations of the Leggett-type inequalities for the tripartite GHZ state. The maximal violation is $2(\sqrt{5}+1)$.}
\label{fig:leggett_violation_ghz}
\end{figure}

The maximal violation is (Fig.~\ref {fig:leggett_violation_ghz}):
\begin{align}
\mathcal{L}_{\pm}^{\mathrm{GHZ}} = 2(\sqrt{5} + 1) \approx 6.472 > 6,
\end{align}
achieved at $\theta = \pi - 2\arctan 2$ for $\mathcal{L}_{+}$ and $\theta = 2\arctan 2$ for $\mathcal{L}_{-}$. 

\section{Conclusions}

We have derived rigorous Leggett-type inequalities for arbitrary $N$-partite systems. The derivation yields a fundamental double inequality that is universally valid, from which the Leggett-type bounds follow under specific measurement settings. For the tripartite GHZ state, we demonstrated quantum violations with maximal violation $2(\sqrt{5}+1)$, which is larger than previous results. The correlation function for the GHZ state behaves differently for odd and even numbers of particles, which may lead to different violation patterns. Our results show that nonlocal realism in multipartite systems is experimentally testable.

\section*{Acknowledgements}
This work was supported in part by the National Natural Science Foundation of China (NSFC) under Grants No.~11975236 and No.~11635009.

\bibliographystyle{unsrt}
\bibliography{NLRref}

\end{document}


\doublespacing

\title{Supplementary Material for ``Leggett-type inequalities for testing nonlocal realism in multipartite systems''}

\author{
Abdul Sattar Khan$^{1,2,3}$, Ma-Cheng Yang$^{1}$, and Cong-Feng Qiao$^{1,4,*}$ \\ [0.2cm]
\footnotesize{$^{1}$School of Physical Sciences, University of Chinese Academy of Sciences, Beijing 100049, China}\\
\footnotesize{$^{2}$School of Physics, Shandong University, Jinan, China}\\
\footnotesize{$^{3}$Wilczek Quantum Center, School of Physics and Astronomy, Shanghai Jiao Tong University, Shanghai 200240, China}\\
\footnotesize{$^{4}$Key Laboratory of Vacuum Physics, CAS, Beijing 100049, China}
}
\date{}

\maketitle

\appendix
\section{Derivation of $N$-partite Leggett-type Inequalities}

We formulate Leggett's nonlocal model with the following photon source. Suppose the source emits $N$ photons with definite polarizations $\vec{u}_1,\ldots,\vec{u}_N$ to $A_1,\ldots,A_N$. The joint probability distribution is:
\begin{align}
\mathcal{P}(x|\mathcal{X}) = \int d\lambda\,\rho(\lambda)\mathcal{P}_{\lambda}(x|\mathcal{X}),
\end{align}
with
\begin{align}
\mathcal{P}_{\lambda}(x|\mathcal{X}) &= \frac{1}{2^N}\Bigg[1 + \sum_{i=1}^{N}x_i\mathcal{C}_{\lambda}^{(i)}(\mathcal{X}) + \sum_{i_1<i_2}x_{i_1}x_{i_2}\mathcal{C}_{\lambda}^{(i_1,i_2)}(\mathcal{X}) \notag \\
&\quad + \cdots + x_1x_2\cdots x_N\mathcal{C}_{\lambda}^{(1,2,\ldots,N)}(\mathcal{X})\Bigg].
\end{align}

The correlation functions are:
\begin{align}
&\mathcal{C}_{\lambda}^{(i)}(\mathcal{X}) = \sum_{x_1,\ldots,x_N} x_i\,\mathcal{P}_{\lambda}(x|\mathcal{X}), \\
&\mathcal{C}_{\lambda}^{(1,\ldots,N)}(\mathcal{X}) = \sum_{x_1,\ldots,x_N} x_1\cdots x_N\,\mathcal{P}_{\lambda}(x|\mathcal{X}), \\
&\mathcal{C}_{\lambda}^{(i_1,\ldots,i_j)}(\mathcal{X}) = \sum_{x_1,\ldots,x_N} x_{i_1}\cdots x_{i_j}\,\mathcal{P}_{\lambda}(x|\mathcal{X}).
\end{align}

The no-signaling condition gives:
\begin{align}
&\mathcal{C}_{\lambda}^{(i)}(\mathcal{X}) = \mathcal{C}_{\lambda}^{(i)}(\mathcal{X}^{(i)}), \\
&\mathcal{C}_{\lambda}^{(i_1,\ldots,i_j)}(\mathcal{X}) = \mathcal{C}_{\lambda}^{(i_1,\ldots,i_j)}(\mathcal{X}^{(i_1)},\ldots,\mathcal{X}^{(i_j)}).
\end{align}

Malus' law requires:
\begin{align}
\mathcal{C}_{\lambda}^{(i)}(\mathcal{X}^{(i)}) = \vec{u}_i\cdot \vec{n}^{(i)}.
\end{align}

From the nonnegativity of probabilities $\mathcal{P}_{\lambda}(x|\mathcal{X}) \geq 0$, we derive:
\begin{align}
\left\{
\begin{aligned}
&|\mathcal{C}_{\lambda}^{(i)}(\mathcal{X}^{(i)}) \pm \mathcal{C}_{\lambda}^{(i_1,\ldots,i_{N-1})}(\mathcal{X})| \leq 1 \pm \mathcal{C}_{\lambda}^{(1,\ldots,N)}(\mathcal{X}), \\
&|\mathcal{C}_{\lambda}^{(i)}(\mathcal{X}^{(i)'}) \pm \mathcal{C}_{\lambda}^{(i_1,\ldots,i_{N-1})}(\mathcal{X}')| \leq 1 \pm \mathcal{C}_{\lambda}^{(1,\ldots,N)}(\mathcal{X}').
\end{aligned}
\right.
\label{eq:core}
\end{align}

This can also be obtained algebraically. If $x,y \in \{\pm1\}$, then $|x \pm y| \mp xy = 1$. Taking $x = x_1$ and $y = x_2x_3\cdots x_N$, multiplying by $\mathcal{P}_{\lambda}(x|\mathcal{X})$, and summing gives Eq.~(1) \cite{Deng2011}.

For two measurement settings, define:
\begin{align}
\gamma_{\pm}^{(i)}(\lambda) &= \mathcal{C}_{\lambda}^{(i)}(\mathcal{X}^{(i)}) \pm \mathcal{C}_{\lambda}^{(i)}(\mathcal{X}^{(i)'}), \\
\alpha_{\pm}(\lambda) &= \mathcal{C}_{\lambda}^{(i_1,\ldots,i_{N-1})}(\mathcal{X}) \pm \mathcal{C}_{\lambda}^{(i_1,\ldots,i_{N-1})}(\mathcal{X}'), \\
\beta(\lambda) &= \mathcal{C}_{\lambda}^{(1,\ldots,N)}(\mathcal{X}) + \mathcal{C}_{\lambda}^{(1,\ldots,N)}(\mathcal{X}').
\end{align}

From Eq.~(1), adding the positive inequalities gives:
\begin{align}
\gamma_{\pm}^{(i)}(\lambda) \leq 2 \mp [\alpha_{\pm}(\lambda) - \beta(\lambda)].
\label{eq:20}
\end{align}

Adding the negative inequalities gives:
\begin{align}
-\gamma_{\pm}^{(i)}(\lambda) \leq 2 \pm [\alpha_{\pm}(\lambda) + \beta(\lambda)].
\label{eq:21}
\end{align}

Equations (13) and (14) are the fundamental inequalities derived from Eq. (1). They can be rewritten as:
\begin{align}
\gamma_{\pm}^{(i)}(\lambda) &\leq 2 - |\alpha_{\pm}(\lambda) - \beta(\lambda)|, \\
-\gamma_{\pm}^{(i)}(\lambda) &\leq 2 - |\alpha_{\pm}(\lambda) + \beta(\lambda)|.
\end{align}

Combining them yields the universally valid double inequality:
\begin{align}
-2 + |\alpha_{\pm}(\lambda) + \beta(\lambda)| \leq \gamma_{\pm}^{(i)}(\lambda) \leq 2 - |\alpha_{\pm}(\lambda) - \beta(\lambda)|.
\label{eq:double}
\end{align}

\medskip
\noindent\textbf{A note on the derivation:} The double inequality in Eq.~(\ref{eq:double}) is the most general and rigorous form of our Leggett-type constraint. It is always valid and follows directly from the axioms. 

A common but incorrect step is to attempt to combine the two inequalities into a single $ \max $ or $ \min $ bound of the form:
\begin{equation}
|\gamma| \leq 2 - \max(A,B).
\end{equation}
This step is not valid in general. For example, if $\alpha = 2$ and $\beta = -1$, then $A = 3$ and $B = 1$, and the bound gives $|\gamma| \leq -1$, which is impossible. The double inequality in Eq.~(\ref{eq:double}) correctly captures the full constraint.

In the specific case where the measurement settings satisfy $(\alpha_{\pm})_k = 0$, the double inequality reduces to the simpler form used in the main text. All violation results reported in this work are obtained under these specific measurement settings and remain valid.
\medskip

Integrating Eq.~(\ref{eq:double}) over $\lambda$ and using $\int d\lambda\,\rho(\lambda) = 1$, we obtain:
\begin{align}
-2 + B \leq \gamma_{\pm}^{(i)} \leq 2 - A,
\end{align}
where
\begin{align}
A &= |\alpha_{\pm} - \beta|, \\
B &= |\alpha_{\pm} + \beta|, \\
\gamma_{\pm}^{(i)} &= \int d\lambda\,\rho(\lambda)\gamma_{\pm}^{(i)}(\lambda), \\
\alpha_{\pm} &= \int d\lambda\,\rho(\lambda)\alpha_{\pm}(\lambda), \\
\beta &= \int d\lambda\,\rho(\lambda)\beta(\lambda).
\end{align}

For three $2N$-tuples of settings with the same angle $\theta$, using $\sum_{k=1}^{3}|\vec{u}_i\cdot \vec{e}_k| \geq 1$, we obtain:
\begin{subequations}
\begin{align}
\sum_{k=1}^{3} |\gamma_{\pm,k}^{(i)}| + \sum_{k=1}^{3} \max\left(|(\alpha_{\pm})_k - \beta_k|, |(\alpha_{\pm})_k + \beta_k|\right) \leq 6.
\label{eq:final}
\end{align}
\end{subequations}

This is the main inequality. Under measurement settings where $(\alpha_{\pm})_k = 0$, it reduces to:
\begin{subequations}
\begin{align}
\mathcal{L}_{+} &\equiv \sum_{k=1}^{3}|\beta_k| + 2|\cos(\theta/2)| \leq 6, \\
\mathcal{L}_{-} &\equiv \sum_{k=1}^{3}|\beta_k| + 2|\sin(\theta/2)| \leq 6.
\end{align}
\end{subequations}

These are the Leggett-type inequalities used in the main text \cite{Branciard2008}.

\section{Correlation Functions for the GHZ State}

The density matrix for the $N$-partite GHZ state is:
\begin{align}
\rho = |\mathrm{GHZ}\rangle\langle\mathrm{GHZ}| = \frac{1}{2}
\begin{pmatrix}
1 & 0 & \cdots & 0 & 1 \\
0 & 0 & \cdots & 0 & 0 \\
\vdots & \vdots & \ddots & \vdots & \vdots \\
0 & 0 & \cdots & 0 & 0 \\
1 & 0 & \cdots & 0 & 1
\end{pmatrix}_{2^N \otimes 2^N}.
\end{align}

The correlation function is:
\begin{align}
\mathcal{C}(\mathcal{X}) &= \tr\left[\rho\left(\bigotimes_{j=1}^{N}\vec{\sigma}\cdot\vec{n}^{(j)}\right)\right] \\
&= \begin{cases}
\Re\left[\prod_{j=1}^{N}(n_1^{(j)} + i n_2^{(j)})\right], & N \text{ odd}, \\[6pt]
\prod_{j=1}^{N}n_3^{(j)} + \Re\left[\prod_{j=1}^{N}(n_1^{(j)} + i n_2^{(j)})\right], & N \text{ even}.
\end{cases}
\end{align}

In spherical coordinates:
\begin{align}
\mathcal{C}(\mathcal{X}) = \begin{cases}
\cos\varphi \prod_{j=1}^{N}\sin\vartheta_j, & N \text{ odd}, \\[6pt]
\prod_{j=1}^{N}\cos\vartheta_j + \cos\varphi \prod_{j=1}^{N}\sin\vartheta_j, & N \text{ even},
\end{cases}
\end{align}
where $\varphi = \sum_{j=1}^{N}\varphi_j$.

For measurements in the $x$-$y$ plane:
\begin{align}
\mathcal{C}(\mathcal{X}) = \cos\left(\sum_{j=1}^{N}\varphi_j\right).
\end{align}

The reduced correlation function for $N-1$ particles is:
\begin{align}
\mathcal{C}(\mathcal{X}_{k}^{(i_1)},\ldots,\mathcal{X}_{k}^{(i_{N-1})}) = \begin{cases}
\prod_{j=1}^{N-1}n_3^{(j)}, & N \text{ odd}, \\[4pt]
0, & N \text{ even}.
\end{cases}
\end{align}

\section{Quantum Predictions for the Tripartite GHZ State}

With the measurement settings in Fig.~1 of the main text, the quantum predictions are:
\begin{subequations}
\begin{align}
\mathcal{L}_{+}^{\mathrm{GHZ}} &= 2\sqrt{5}\sin(\theta_0 + \theta/2) + 2\left|\cos\frac{\theta + \phi + \varphi}{2}\right|, \\
\mathcal{L}_{-}^{\mathrm{GHZ}} &= 2\sqrt{5}\cos(\theta_0 - \theta/2) + 2\left|\cos\frac{\theta + \phi + \varphi}{2}\right|,
\end{align}
\end{subequations}
where $\theta_0 = \arctan 2$.

The maximal violation is:
\begin{align}
\mathcal{L}_{\pm}^{\mathrm{GHZ}} = 2(\sqrt{5} + 1) \approx 6.472 > 6,
\end{align}
achieved at $\theta = \pi - 2\arctan 2$ for $\mathcal{L}_{+}$ and $\theta = 2\arctan 2$ for $\mathcal{L}_{-}$.

\begin{figure}[H]\centering
\scalebox{0.60}{\includegraphics{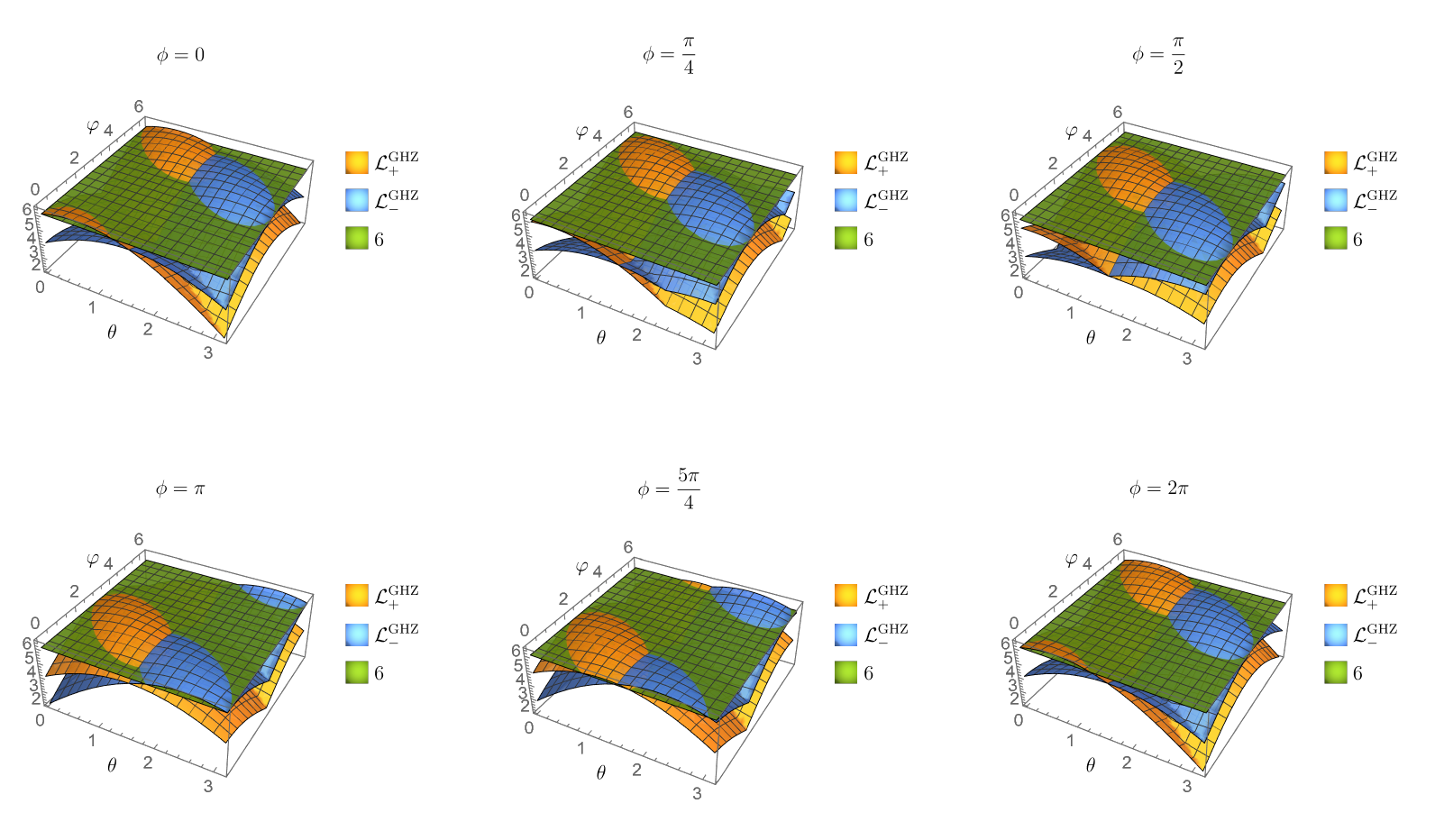}}
\caption{Quantum violations for the tripartite GHZ state.}
\label{fig:supp_leggett_violation_ghz}
\end{figure}

\bibliographystyle{unsrt}
\bibliography{SMref}